# Evolution: Google vs. DRIS

Wang Liang[1], Guo Yiping[2], Fang Ming[1]

1 (Department of Control Science and Control Engineer, Huazhong University of Science and Technology, WuHan, 430074 P.R.China)

2(Library of Huazhong University of Science and Technology, WuHan 430074 P.R.China)

E-mail:wangliang_f@yahoo.com

**Abstracts**: **This paper gives an absolute new search system that builds the information retrieval infrastructure for Internet. Now most search engine companies are mainly concerned with how to make profit from company users by advertisement and ranking prominence, but never consider what its real customers will feel. Few web search engines can sell billions dollars just at the cost of inconvenience of most Internet users, but not its high quality of search service. When we have to bear the bothersome advertisements in the awful results and have no choices, Internet as the kind of public good will surely be undermined. If current Internet can't fully ensure our right to know, it may need some sound improvements or a revolution.**

## 1 Introduction

For many scientists, finding the information they really want on Internet is still a hit-and-miss affair. There are billions of web pages, millions of databases and many other kinds of information resources on Internet, but there still no an efficient method to manage and utilize these exponentially growing information. As the "information management system" of Internet, any current search engines can't continue to index close to the entire Web as it grows (1). "Too much information means no information". Obtaining the precise and comprehensive information from all these resources is becoming a kind of acrobatics.

On the other hand, Internet belongs to every of us, but Google is its god. We normally obtain information from search engine. But it administrates what you could know and what you couldn't reach to a large extent. If all the information is in the charge of a small group, Internet will surely become its own vault (2).

Here we show our work, an absolutely new kind of search system, DRIS (Domain resources integrate system), which is built as an infrastructure of Internet. The basic idea of DRIS is that search should be the internal function of Internet and everyone should have his own personal intelligent search engine. This system will absolutely transform the basic idea of search engine and return dominion of Internet to every of us from search engine company. You will get what you really want from your own search engine: precise, comprehensive, fresh information. We first briefly introduce three kinds of current information retrieval system, and then describe the basic content of DRIS.

## 2 Three kinds of search system

Besides web search engines, we can also obtain information from other kinds of information retrieval system such as FTP, P2P and many databases in library. All this information retrieval systems can be divided into three kinds by architecture.

1 centralized search system (Fig.1). This information retrieval system has its own data collecting mechanism, and all the data are stored and indexed in a conventional database. Most of web search engines belong to this kind.

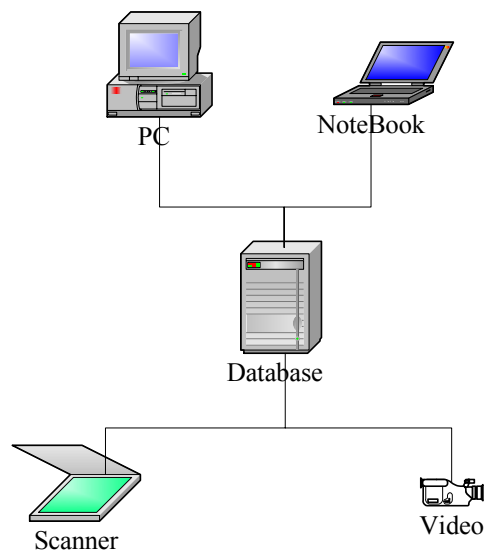

Figure1 search engine based on database system

2 metadata harvest system (Fig.2). When we need to integrate different kinds of information resource, we can harvest the metadata from sub databases and build a union metadata database to provide the combined search function. Normally Metadata is much smaller than data itself, for example, a video record is 1GB but its metadata may be only 1kB. Some systems based on OAI like NSDL just apply this method.

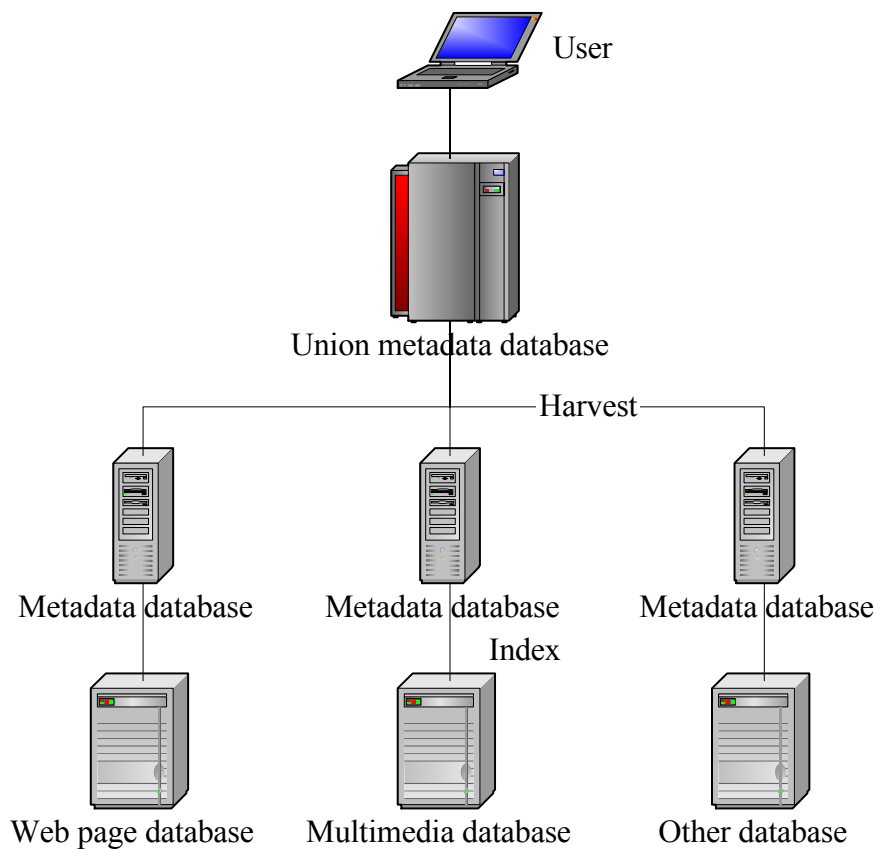

Figure 2 in information retrieval system based on metadata harvest architecture, the metadata of sub databases will be timely downloaded by a harvest mechanism. Users can obtain different kinds of information from a union metadata database.

3 Distributed search system (Fig.3). If the data sources are large enough and the aggregation of Metadata is also too huge to store in a system, we can use distributed frame to integrate all the resources. Distributed search system hasn't any actual data resources, but an index that descript the search interface and other characters of its sub data resources. Different resources will cooperate to carry out the user's query. InfoBus (3) is this kind of search system.

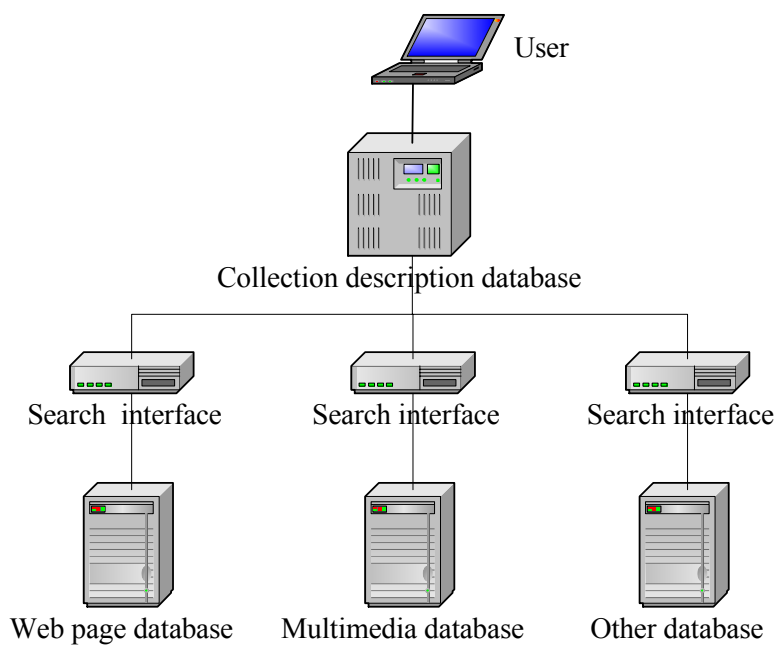

Figure 3 distributed information retrieval system has no his own actual record database. When receiving a query from a user, it will instantly obtain the records through the search interfaces provided by sub databases.

## 3 how to select correct architecture

There are two main characters to determine the architecture of a information retrieval system, the size and diversity of data source. We can select appreciate appropriate according to the principle shown in Fig.4.

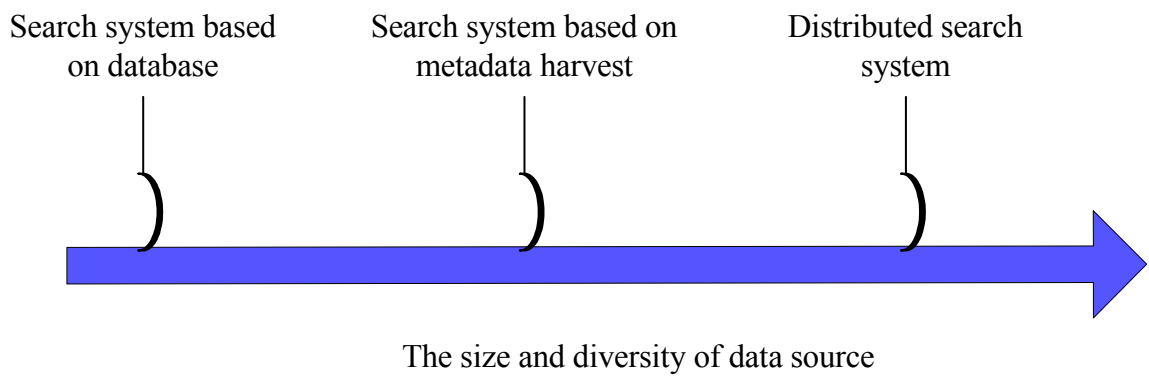

Figure4 the principle about how to select correct architecture

Internet is the biggest database in this world. We can't use a single kind of search system to manage this database. But most web search engines only apply the centralized structure. Some limitations of current web search system have been very obvious. So how to correctly apply three existing search architecture to build a system that can manage all the information on Internet is an urgent problem. We just proposed such a system, Domain resource integrated system (DRIS), and are building the testbed in CERNET (China Education and Research Network).

## 4 What is DRIS?

1 Architecture

DRIS is hierarchical information manage system, whose architecture is as same as that of DNS. Domain (defined in DNS) is the basic cell of DRIS. In every domain, a central server is built to manage information under this domain (4). The architecture of DRIS is shown in Fig.5.The node in the third layer corresponds to a third level domain like a university, and the second layer may be the sub network of one country like domain of "edu.cn". The node in top layer normally is a country. Every node of DRIS is an integrated search engine, but has different search scopes.

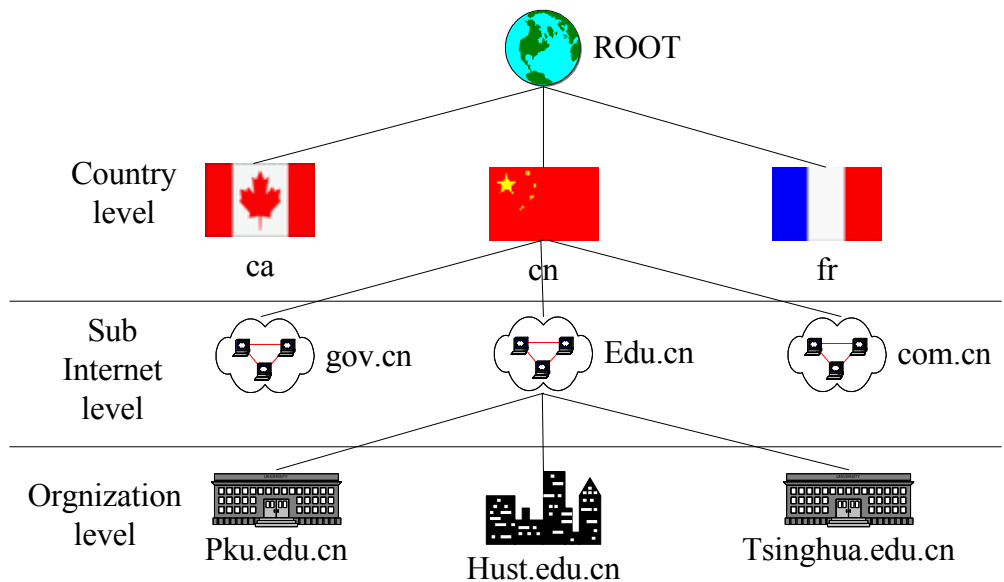

Figure 5 the architecture of DRIS is as same as that of DNS. DRIS is a hierarchical distributed information retrieval system. Every node of DRIS will provide standard search service.

## 2 function structure

DRIS appropriately adopts three kinds of search system in three layers.

In the third layer, centralized database system is applied to build search system. Node in this layer is always limited in a university or a company, and the size of its data source is very small. For example, the number of web pages in a university normally isn't more than one million, and most of current database systems can easily manage this kind of data source.

In the second layer, we adopt the metadata harvest system to build the information retrieval system. As a web search engine of node "edu.cn", we will integrate the all the web pages in all universities in China. Merely harvesting the metadata from thousands of DRIS servers in the third layer will be more efficient than directly downloading all the web pages from millions of web servers. The other kinds of resources can also be integrated by this means.

In the first layer, Distributed information retrieval system is applied to integrate the information resource in the scope of a country. In this layer, only the collection description data of resources is stored in database of DRIS. All the resources will cooperate to finish a user's query under the directing of DRIS. The function structure of DRIS is shown in Fig.6.

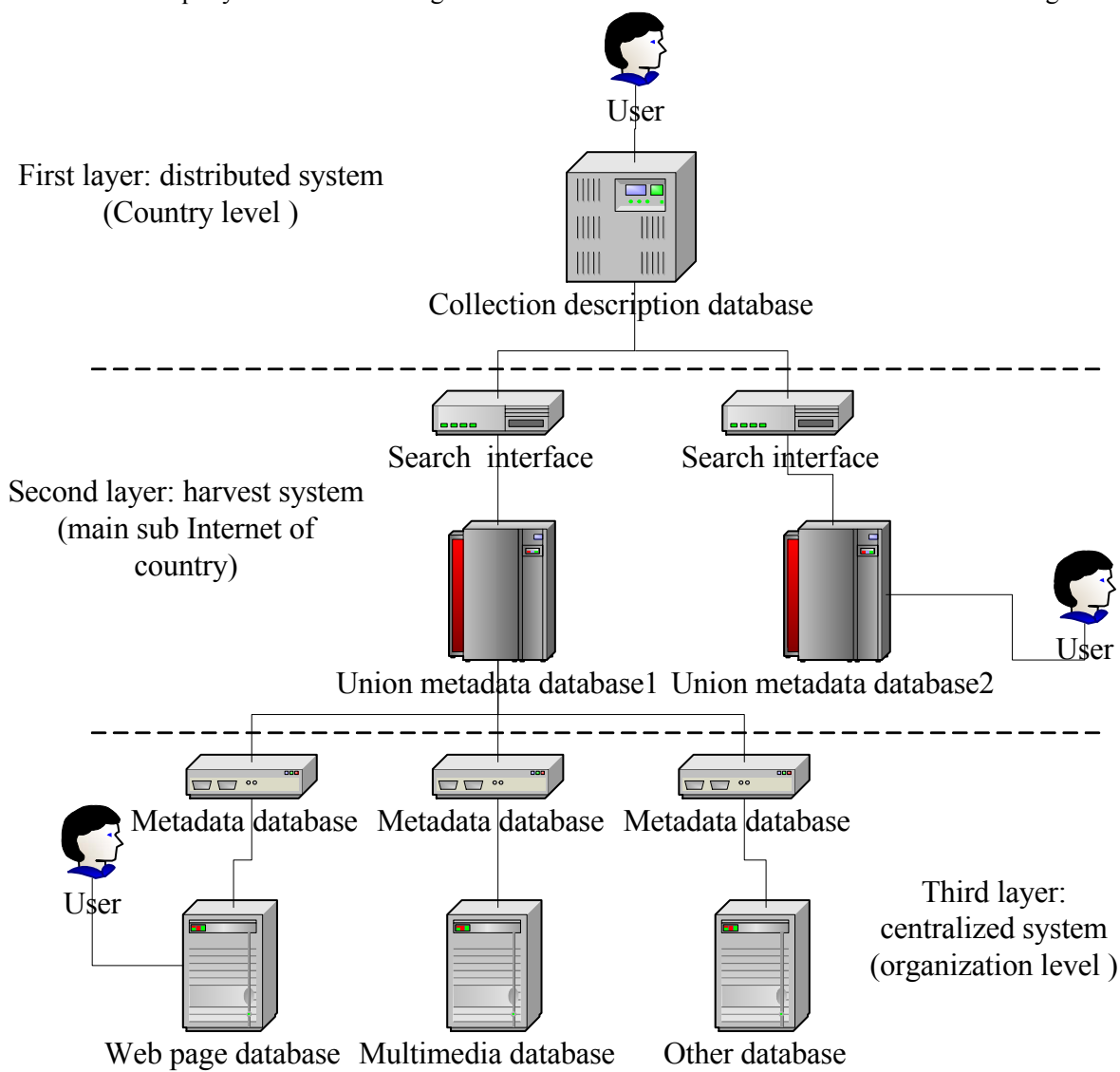

Figure 6 DRIS suitably applies three kinds of information retrieval system to build the infrastructure of information retrieval for Internet. Organization lever- centralized system. Sub Internet level – metadata system. Country level- distributed system. DRIS can be treated as the management system of the biggest database in this world-Internet.

The size of data source markedly increases from the third layer to first layer. In appropriate scope (three levels domain), applying appropriate information retrieval architecture, and building appropriate information management frame for the whole Internet is the basic principle of DRIS.

The differences between DRIS and Google are detailed as follows.

1 Internal and external. DRIS is the internal component of Internet, but the search engine is the external merchandise of a company. DRIS is built in different local networks, so it can fully use its high communication speed. On the contrary, Google may have to download the pages from a distance server. DRIS is managed by its users, just like management method of DNS. Every organization is its customer and also its builder. Decentralized management is much more effective than centralized administration in a large scale system. As the infrastructure of Internet, DRIS can almost cover all the information resources on Internet in theory. At the same time, its update interval can be only one day.

2 Opening and private. Search engines belong to corresponding companies, so they have to obtain profit from advertisement and ranking to maintain their subsistence. But DRIS is opening system, which needn't any profits from its users and of course need not any advertisements. Furthermore, In DRIS, every node will provide the standard search interface for anyone. So many intelligent personal search systems can use DRIS as their data source and provide high quality of search service. On the contrary, most search engines can only provide the same results for the same query words submitted by anyone.

**5 The secret of DRIS**

DRIS will enormously improve the performance of Internet search engine in recency, coverage and so on, but this can't ensure the establishment of DRIS. The secret of DRIS will lies that it can balance the benefit and obligation between different organizations. Just in our testbed, education network, only few universities have the web search engine for the school network. Every university has also many characteristic information resources such as Ftp, BBS and special databases in library, but almost no a university have union search system that can efficient integrate all these resources, say nothing of sharing these resources with each other. These all bring the request for creating the underlying structure of DRIS. Solving some urgent problems of his own and then benefiting others may be the real guarantee of the success of DRIS. Moreover, the implementations of IPV6 will great prompt the information explosion, and also bring the requests for building an infrastructure of information management.